\documentclass[prl,twocolumn,showpacs,superscriptaddress,amsmath,amssymb]{revtex4}
\usepackage{amsmath}
\newif\ifpdf
    \ifx\pdfoutput\undefined
    \pdffalse % we are not running PDFLaTeX
    \else
    \pdfoutput=1 % we are running PDFLaTeX
    \pdftrue
    \fi

    \ifpdf
    \usepackage[pdftex]{graphicx}
    \else
    \usepackage{graphicx}
    \fi

\begin{document}

\title{Optimizing photon indistinguishability
in the emission from incoherently-excited semiconductor quantum
dots}

\author{F. Troiani}
\email[]{filippo.troiani@uam.es} \affiliation{Departamento de
Fisica Teorica de la Materia Condensada, Universidad Autonoma de
Madrid, 28049 Madrid, Spain}
\author{J. I. Perea}
\affiliation{Departamento de Fisica Teorica de la Materia
Condensada, Universidad Autonoma de Madrid, 28049 Madrid, Spain}
\author{C. Tejedor}
\affiliation{Departamento de Fisica Teorica de la Materia
Condensada, Universidad Autonoma de Madrid, 28049 Madrid, Spain}

\date{\today}

\begin{abstract}

Most optical quantum devices require deterministic single-photon
emitters. Schemes so far demonstrated in the solid state imply an
energy relaxation which tends to spoil the coherent nature of the
time evolution, and with it the photon indistinguishability. We
focus our theoretical investigation on semiconductor quantum dots
embedded in microcavities. Simple and general relations are
identified between the photon indistinguishability and the
collection efficiency. The identification of the key parameters
and of their interplay provides clear indications for the device
optimization.

\end{abstract}


\maketitle

Deterministic solid-state single-photon sources (S4Ps) are
fundamental building blocks of potential quantum devices in the
areas of quantum communication, cryptography~\cite{nielsen}, and
computation~\cite{knill:01}. Each of these applications sets a
number of stringent requirements for the S4P to fulfill, including
an efficient collection of the emitted light, and highly
non-classical (sub-Poissonian) photon correlations. Besides, the
eventual manipulation of the quantum information relies on
two-photon interference, and thus requires the photons to be
prepared in a given (pure) quantum state: this typically calls for
a suppression of the de-cohering interactions with the solid-state
environment. A number of S4Ps have been demonstrated in recent
years, based on the use of either vacancies~\cite{exp_vac},
molecules~\cite{exp_mol}, or mesoscopic
heterostructures~\cite{santo:02,exp_dots}. The latter,
specifically consisting in semiconductor quantum dots (QDs), are
particularly attractive due to the increasing degree of accuracy
with which they can be embedded in (and coupled to) different
kinds of optical microcavities (MCs)~\cite{badol:05}.

The requirement of generating indistinguishable photons translates into that of
driving the emitter's time evolution in a coherent fashion from the ground- to
the radiating-state, within timescales shorter than those the decoherence acts
on.
In this respect, coherent-carrier control in semiconductor nanostructures, based
on the use of ultrafast laser pulses, was proven to be a powerful
technique~\cite{car_con}.
The excitation of the system by optical means, however, brings about the
practical problem of discriminating the outcoming photon from the incoming
radiation.
While more sophisticated approaches have been theoretically
proposed~\cite{theory,kiraz:04},
the schemes so far applied at an experiment level rely on the energy relaxation
the quantum emitter undergoes between the excitation and the emission processes.
The incoherent nature of such evolution, and the resulting classical uncertainty
on the starting time of the photon-emission process (time-jitter), could render
such schemes inherently inadequate to the implementation of a
S4P~\cite{kiraz:04}.
It is the purpose of the present paper to gain a deeper insight into the above
issue, and to show that, to a large extent, a full exploitation of the system
engineering allows to circumvent such limitations.

\begin{figure}[tbp]
\begin{center}
\includegraphics[width=0.75\columnwidth]{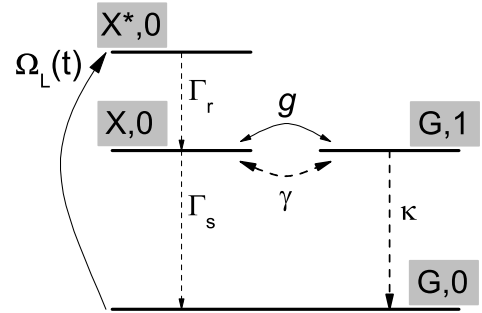}
\caption{Schematics of the dot-cavity system's relevant states: $
 |G,0\rangle, |G,1  \rangle,
 |X,0\rangle, |X^*,0\rangle $.
The occupation of the remaining states is negligible throughout
system's time evolution.
In all the calculations presented below, we consider the case of a QD
resonantly excited from
$|G\rangle$ to $|X^*\rangle$ by means of a gaussian laser pulse
$\Omega_L (t) = \Omega_0 \exp [(t-t_0)^2 /2\sigma^2] $,
with $\sigma = 1.1$~ps and $\Omega_0=0.75$~meV.}
\label{fig1}
\end{center}
\end{figure}
The physical system we shall refer to in the following consists in
a single semiconductor QD, resonantly coupled to an optical MC.
The discrete nature of the dot spectrum and the large interlevel
spacings characterizing self-assembled QDs allow the selective
addressing of its interband transitions by means of laser-pulses
at the ps timescale~\cite{zrenner}. As a consequence, the system
can be substantially prevented from being multiply excited, and
its state vector effectively confined within a reduced portion of
the Hilbert space. We accordingly restrict ourselves to the a few
QD states: the ground (or vacuum) state $|G\rangle$, the lowest
exciton state $|X\rangle$, and the first excited among the
optically active ones $|X^*\rangle$. Besides, we consider a single
cavity mode, with no a priori restriction on its occupation number
$n$. The probability of a multiple-excitation ($n>1$, or $n>0$
with the QD in one of its exciton states) turns out to be
negligible throughout the system's time evolution for the
considered range of physical parameters: the level structure we
shall refer to in the discussion of our results thus reduces to
the one represented in Fig.~\ref{fig1}.

In order to account for the open nature of the dot-cavity system, namely for
its coupling with the phonon and the photon reservoirs, we describe the
time-evolution of the density operator $\rho$ by means of the following
master equation in the Lindblad form:
\begin{eqnarray}
\dot{\rho} =
\frac{i}{\hbar} [\rho , H] +
\frac{\gamma}{2} \sum_{i=1}^3 (2 \mathbb{P}_i\, \rho\, \mathbb{P}_i -
     \mathbb{P}_i\, \rho - \rho\, \mathbb{P}_i ) +
\sum_{\alpha=r,s}\frac{\Gamma_{\alpha}}{2} \times
\nonumber \\
(2 \sigma_{\alpha} \rho \sigma^{\dagger}_{\alpha} -
     \sigma^{\dagger}_{\alpha} \sigma_{\alpha} \rho - \rho
     \sigma^{\dagger}_{\alpha} \sigma_{\alpha}) +
\kappa (2 a \rho a^{\dagger} - a^{\dagger} a \rho - \rho a^{\dagger} a),
\label{eq:master}
\end{eqnarray}
where
$\sigma_s = | G   \rangle\langle X   |$,
$\sigma_r = | X   \rangle\langle X^* |$,
$ \mathbb{P}_1 = | G   \rangle\langle G   | $,
$ \mathbb{P}_2 = | X   \rangle\langle X   | $,
and
$ \mathbb{P}_3 = | X^* \rangle\langle X^* | $.
The constant and time-dependent components of the Hamiltonian $H$, given by
$H_{DC} =
%\omega_{X  } | X   \rangle \langle X   | +
%\omega_{X^*} | X^* \rangle \langle X^* | +
\hbar g\, (\sigma_s a^{\dagger} + a \sigma^{\dagger}_s)$
and
$H_L = - (\hbar / 2)\, \Omega_L (t)\, (\sigma_r + \sigma^{\dagger}_r)$,
account for the coherent couplings of the dot with the MC and with the driving
laser, respectively.
Further details on the model can be found in Ref.~\cite{perea:04}.

\begin{figure}[tbp]
\begin{center}
\includegraphics[width=\columnwidth]{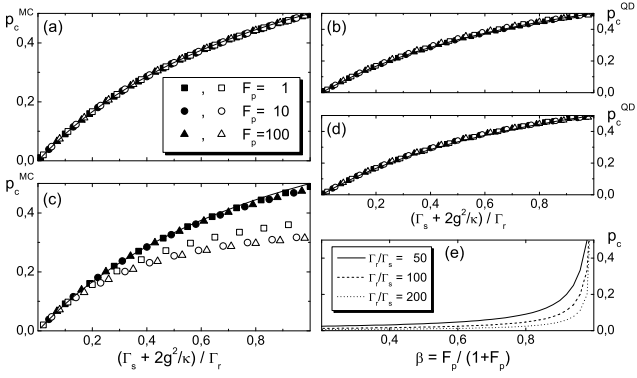}
\caption{ (a-d) Distinguishability of the photons emitted by the microcavity
($p_c^{MC}$) and by the quantum dot ($p_c^{QD}$) as a function of the overall
rate emission $R \equiv \Gamma_s + 2 g^2 /\kappa$ normalized to $\Gamma_r$, in
the absence of dephasing ($\gamma=0$).
Different symbols stand for $F_p=1$ (squares), $F_p=10$ (circles) and $F_p=100$
(triangles).
Solid lines depict the curve $p_c=R/(R+\Gamma _r)$.
In each plot, $\Gamma_s$ and either $g$ [panels (c,d)] or $\kappa$ (a,b) are
varied in such a way that $\Gamma_s =R/(F_p+1)$ and $2g^2 / \kappa = R F_p /
(F_p+1)$.
The fixed parameters are: (a,b) $\kappa =10\ {\rm ps}^{-1}$, $\Gamma_r = 0.2
\ {\rm ps}^{-1}$ (black symbols) and $0.05$ (white symbols); (c,d) $\Gamma_r =
0.1\ {\rm ps}^{-1}$, $g=0.2\ {\rm meV}$ (black) and $0.05$ (white).
(e) Empirical relation between the collection efficiency $\beta$ and $p_c$
for different values of $\Gamma_r / \Gamma_s$.}
\label{fig2}
\end{center}
\end{figure}
Interference represents a natural means to measure the overlap
between the quantum states of two photons consecutively emitted
(one per driving pulse) by the single-photon source. Within a
Hong-Ou-Mandel type experiment~\cite{hong:87,santo:02}, the
indistinguishability is reflected by the bunching behavior of the
two photons entering the two input channels of a balanced
beam-splitter. More specifically, a perfect overlap between the
photons' state vectors completely suppresses the probability of
measuring one photon in each of the two output channels. The
experimentally accessible quantity accounting for such coincidence
probability can be obtained from the emitter's first-order
coherence functions~\cite{kiraz:04}:
\begin{equation}
p_c^{\chi} =\frac{
\int_0^T dt \int_0^{T-t} d\tau\, [\,
G^{(1)}_{\chi} (t)\, G^{(1)}_{\chi} (t+\tau) -
| G^{(1)}_{\chi} (t,\tau) |^2]\, }{
\int_0^T dt\,
G^{(1)}_{\chi} (t) \int_0^{T-t} d\tau\, G^{(1)}_{\chi} (t+\tau)}  ,
\label{eq:indist}
\end{equation}
where $G^{(1)}_{QD} (t,\tau) \equiv \langle\,
\hat{\sigma}^{\dagger}_s (t) \, \hat{\sigma}_s (t+\tau)\, \rangle
$, $G^{(1)}_{MC} (t,\tau) \equiv \langle\, \hat{a}^{\dagger} (t)
\, \hat{a} (t+\tau)\, \rangle $, $ G^{(1)}_{\chi} (t) \equiv
G^{(1)}_{\chi} (t,0)$ ($\chi = MC,QD$); $T$ is the time interval
between two consecutive laser pulses, which we assume larger than
that the system requires to emit the photon and relax back to its
ground state.

For the sake of clarity, we start by considering the photon properties in the
absence of dephasing ($\gamma =0$).
In Fig.~\ref{fig2} (a-d) we show the coincidence probabilities for the photons
emitted by the QD and by the MC ($p_c^{QD}$ and $p_c^{MC}$, respectively).
Each symbol (square, circle, or triangle) corresponds to a given value of the
Purcell factor, $F_p=2g^2/\kappa \Gamma _s$, i.e. to a specific ratio between
the MC and QD emission rates;
the overall rate emission $R = \Gamma_s +2g^2/\kappa$ increases along the
horizontal axis from $0.01$ to $1$ times $\Gamma_r$, while either $\kappa$ or
$g$ are kept constant [panels (a,b) and (c,d), respectively].
The photon indistinguishability clearly depends on the physical parameters
solely through the overall-emission rate $R$ normalized to $\Gamma_r$,
while it is hardly affected by orders-of-magnitude changes in the Purcell
factor.
Moreover, such dependence is very well approximated by the simple expression
\begin{equation}
p_c \equiv R/(R+\Gamma _r)
\label{distinguishability}
\end{equation}
(solid curve in each of the panels), which is generally valid for
both $\chi=QD$ and $\chi=MC$. In fact,
Eq.~\ref{distinguishability} can be analytically proven to be the
exact solution for a simplified, three-level version of the system
sketched in Fig.~\ref{fig1}, in the instantaneous-excitation
limit~\cite{analytical}; quite generally, it holds as long as the
only effect of the MC on the system's dynamics is that of
enhancing the QD emission rate by a factor $(1+F_p)$, while it
overestimates $p_c^{MC}$ for $g \lesssim \Gamma_r , R$ [see panel
(c) and the discussion below].

Within a semiconductor-based S4P, the role of the MC is also that of
meaningfully enhancing the collection efficiency~\cite{gerar:02}.
In fact, the MC photons are typically emitted in one or few well defined
directions, while the emission from the QD has a more isotropic  character:
the photon-loss probability can thus be approximately identified with that of
the radiation being emitted directly by the QD into the leaky modes.
Hereafter we concentrate in the MC photon emission, for which we assume a
maximum collection efficiency.
The calculated fraction of photons emitted from the cavity is
$\beta = N_{MC} / (N_{MC} + N_{QD})$,
being
$ N_{MC} = 2\kappa\int_0^T dt\, \langle a^{\dagger} (t) \, a (t) \rangle$,
and
$ N_{QD} = \Gamma_s\int_0^T dt\, \langle \sigma^{\dagger}_s (t) \, \sigma_s
(t) \rangle $.
Within the considered range of physical parameters, the collection efficiency
is readily expressed as a function of the Purcell factor~\cite{gerar:02}:
$\beta = F_p / (F_p+1) = 1 - \Gamma _s / R$.
The existence of simple and general expressions for both $p_c^{\chi}$ and
$\beta$ allows to establish between the two a relation which assignes to
the ratio $\Gamma_r / \Gamma_s$ a key role in the performance of the device:
\begin{eqnarray}
p_c= [ 1+\frac{\Gamma _r}{\Gamma _s} (1- \beta) ]^{-1}.
\label{distinguishability2}
\end{eqnarray}
While the degree of photon indistinguishability monotonically decreases with
increasing collection efficiency [Fig.~\ref{fig2} (e)]~\cite{kiraz:04}, the
possibility of contextually achieving for both $\beta$ and $1-p_c$ values
close to 1 sensitively depends on the ratio $\Gamma_r / \Gamma_s$, and calls
for its maximization.
As discussed below, positive or negative corrections to such estimate of the
coincidence probability may arise respectively from the presence of dephasing
and from a proper weakening of the dot-cavity coupling.

\begin{figure}[tbp]
\begin{center}
\includegraphics[width=\columnwidth]{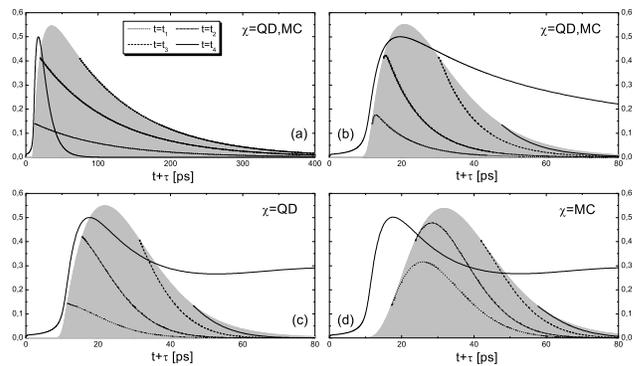}
\caption{Degree of purity and coherence of the dot-cavity system as a function of
time.
The purity (thick curves) is quantified by
$ f_{\tilde{\rho}}(t+\tau) \equiv 1 - {\rm Tr} [\tilde{\rho}^2 (t+\tau)] /
{\rm Tr} [\tilde{\rho}(t+\tau)]^2 $,
calculated within the subspace
$ \{| G,1 \rangle, |X,0 \rangle, |X^*,0 \rangle \} $.
The coherence (narrow curves) is given by
$ | G^{(1)}_{\chi} (t,t+\tau) |^2 \, / \, G^{(1)}_{\chi} (t) $,
corresponding to different initial times $t=t_i$, and enveloped by
$ G^{(1)}_{\chi} (t+\tau) $
(shaded grey); the times $t_i$ are such that
$ G^{(1)}_{\chi} (t_{2,3}) = 3\, G^{(1)}_{\chi} (t_{1,4}) =
0.75\, \max \{ G^{(1)}_{\chi} (t) \}$.
The labels on the vertical axes refer to $ f_{\tilde{\rho}}$, while the other
curves are plotted in arbitrary units.
$ R / \Gamma_r $ is $1.0$ in panel (a) and $0.1$ in (b-d); $ g $ is $2.0$~meV in
panels (a,b) and $0.05$ in (c,d); $\Gamma_r = 0.1\ {\rm ps}^{-1}$ in (a-d).}
\label{fig3}
\end{center}
\end{figure}

A deeper physical insight into the above behaviors can be gained
by looking at the time evolution of the populations
$\rho_{\alpha,\alpha}$ and of the correlation functions. The
narrow lines in Fig.~\ref{fig3} correspond to $ | G^{(1)}_{\chi}
(t,t+\tau) |^2 \, / \, G^{(1)}_{\chi} (t) $, plotted as functions
of $ \tau $ and for different values of the initial times $t=t_i$;
the shaded grey areas which envelope them give $ G^{(1)}_{\chi}
(t+\tau) $, and coincide with either $ \rho_{X,0;X,0} $ or $
\rho_{G,1;G,1} $, depending on $\chi$ being equal to QD or MC. In
both cases, being the difference between the two proportional to
the integrand function of Eq.~\ref{eq:indist}, the closer the
black curves are to their grey envelope, the more the photons are
indistinguishable. The thick lines in the plots, instead, quantify
the degree of purity of the dot-cavity system, for they correspond
to the function $ f_{\tilde{\rho}}(t+\tau) \equiv 1 - {\rm Tr}
[\tilde{\rho}^2 (t+\tau)] \, / \, {\rm Tr}
[\tilde{\rho}(t+\tau)]^2 $, with $\tilde{\rho}$ the density matrix
reduced to the subspace $ \{| G,1 \rangle, |X,0 \rangle, |X^*,0
\rangle \} $. Quite generally, a large contribution to the value
of $p_c^{\chi}$ arises from the coherence functions whose initial
time $ t_i $ falls in the raising region of $ G^{(1)}_{\chi} $,
while those with a later $t_i$ tend to approach the envelope. The
intuitive explanation for this feature is that the portion of
photon emitted at any time $t$ is not linearly superimposed to
(and therefore does not interfere with) the contribution arising
from the population of $\rho$ which still has not undergone the
energy relaxation, namely $\rho_{X^*,0;X^*,0} (t)$. The comparison
between the time evolutions corresponding to $R/\Gamma_r = 0.1$
and $R/\Gamma_r = 1.0$ [panels (a) and (b)] clearly shows how a
slow emission strongly reduces the relative importance of the
emission during the rising of $ \rho_{X,0;X,0} $ and $
\rho_{G1;G,1}$, and thus the average mixing of the system during
the radiative process. The very same interpretation applies to the
dependence of $p_c^{MC}$ on the dot-cavity coupling constant $g$
[Fig.~\ref{fig2} (c)]. In fact, while for relatively large values
of $g$, the population of the states $|X,0\rangle$ and
$|G,1\rangle$ occurs simultaneously (Fig.~\ref{fig3}, panel (b)),
for weaker dot-cavity couplings the latter suffers a delay with
respect to the former [panels (c,d)]: the maximum of $
\rho_{G1;G,1} $ is thus displaced with respect to that of
$f_{\tilde{\rho}}$, and the photon emitted by the MC experiences
on average a higher degree of purity with respect to that coming
from the QD. Correspondingly, $p_c^{MC}$ and $p_c^{QD}$ coincide
with $p_c$ for $g>\Gamma_r, R$, while $p_c^{MC} < p_c^{QD} \simeq
p_c$ for $g < \Gamma_r, R$. Thus, against intuition, a relative
strengthening of the QD's incoherent interaction with the phonon
reservoir ($\Gamma_r$) with respect to the coherent dot-cavity
coupling ($g$), results in an increased degree of
indistinguishability of the emitted photon, due to a deviation of
the system's time evolution from that of an effective three-level
system~\cite{analytical}.

\begin{figure}
\includegraphics[width=\columnwidth]{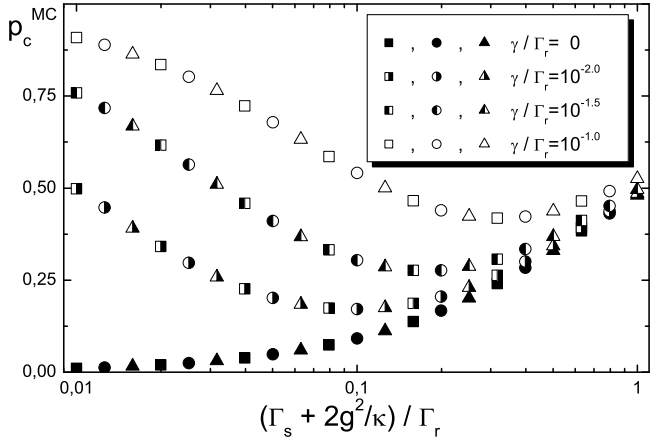}
\caption{Cavity-photon distinguishability as a function the rate emission
$R$, for different values of the dephasing rate $\gamma$ (see the legend) and
of the Purcell factor $F_p$ (squares, circles and triangles correspond
respectively to $F_p=1,10,100$).
Other physical parameters:
$\Gamma_r = 0.1\ {\rm ps}^{-1}$, $g = 0.2\ {\rm meV}$.}
\label{fig4}
\end{figure}

\begin{figure}
\includegraphics[width=\columnwidth]{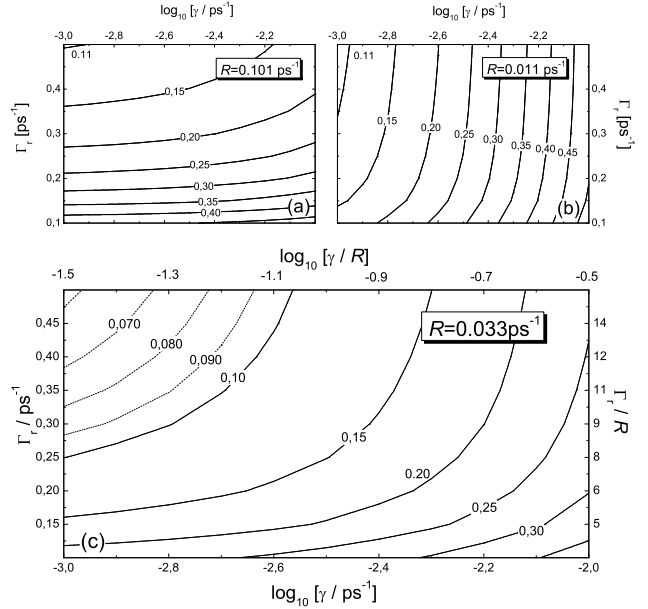}
\caption{Cavity-photon distinguishability as a function of the
non-radiative relaxation and of the dephasing rates, $\Gamma_r$
and $\gamma$, for different values of the rate emission $R$. In
all cases $\Gamma_s = 10^{-3}\ {\rm ps}^{-1}$, while $F_p=100,10,
32$ in panels (a), (b), and (c), respectively. The dot-cavity
coupling constant is $g=0.1\ {\rm meV}$ in (a,b) and $0.05$ in
(c). The behavior of $p_c^{MC}$ for $R=0.033$ and $g=0.1\ {\rm
meV}$ (not shown here) is qualitatively that shown in panel (c),
though the values are on average higher.} \label{fig5}
\end{figure}

Unlike atomic emitters, QDs are permanently coupled to a
solid-state environment, which provides a number of scattering
channels, mainly related to the lattice
vibrations~\cite{rossi,car_con}. These include inelastic processes
(real transitions), which are responsible for the non-radiative
relaxation of the dot from $|X^*\rangle$ to $|X\rangle$, and
elastic ones (virtual transitions), giving rise to the so-called
pure dephasing~\cite{vagov:03}. A detailed investigation of their
physical origin and dependence on the specific features of the QD
is beyond the scope of the present paper; in the following we
restrict ourselves to considering their effects on the properties
of the emitted photons. In Fig.~\ref{fig4} we show the coincidence
probability $p_c^{MC}$ as a function of the rate emission $R$,
normalized to $\Gamma_r$, for different values of $\gamma$ and of
$F_p$. Once again, the key parameter turns out to be $R$, while no
appreciable dependence on the Purcell factor shows up. On the
other hand, a clear competition between $\gamma$ and $R$ emerges:
the largest difference in the coincidence probabilities between
the $\gamma \neq 0$ and the $\gamma=0$ cases occurs in the
slow-emission region; the values of $R$ at which the two values
approach each other grow with increasing dephasing rate. However,
the curves corresponding to different values of $\gamma$ are not
similar, i.e. they don't reveal any simple scaling behavior. We
thus investigate the overall interplay between the relevant
parameters, $\gamma$, $\Gamma_r$, and $R$, by considering the
dependence of $p_c^{MC}$ on the former two for fixed values of the
latter (Fig.~\ref{fig5}).
For $R=0.101\ {\rm ps}^{-1}$ [panel (a)], the photon
indistinguishability is relatively unsensitive to the value of the
dephasing rate, which is graphically rendered by the isolines
being roughly parallel to the $\gamma$ axis (see the left-hand
side of the plot); instead, it meaningfully increases for
increasing $\Gamma_r$. This indicates that already for $R / \gamma
\gtrsim 10$ the requirement that the emission process be faster
than dephasing is substantially satisfied, as suggested by the
results of Fig.~\ref{fig4}. While an order-of-magnitude reduction
of the rate emission [panel (b)] inverts the above situation, the
setting of $R$ to $0.033\ {\rm ps}^{-1}$ [panel (c)] clearly
enhances the area where $1-p_c^{MC}$ exceeds $90$\%. Therefore,
depending on which region of the $(\gamma , \Gamma_r)$ plane can
actually be accessed through the QD engineering and experimental
conditions, different values of the photon-emission rate (and thus
of the cavity-related parameters) maximize the photon
indistinguishability. 
Altogether, coincidence probabilities lower
than $0.1$ can be reached for inelastic scattering rates $\Gamma_r$ roughly
two orders of magnitude larger than the elastic one $\gamma$, and
compromise values of $R$, such that $10\, \gamma \lesssim R
\lesssim 0.1\, \Gamma_r$.

In conclusion, the indistinguishability of the photons emitted by an
incoherently-excited dot-cavity system essentially depends on the relative
efficiency of the system's radiative and non-radiative relaxations, i.e.
on $R/\Gamma_r$.
A simple empirical relation between such ratio and the coincidence
probability is found in the slow-dephasing limit; this in turn allows to
express $p_c$ as a function of $\beta$, and to highlight the key role
played by $\Gamma_r / \Gamma_s$ within the device performance.
The overall interplay between $R$, $\Gamma_r$, and the dephasing rate
$\gamma$ is investigated, providing indications on the optimal parameter
range.

This work was partly supported by the Spanish MCyT under contract
No. MAT2002-00139, CAM under Contract No. 07N/0042/2002,
and the EC within the Research Training Network COLLECT.

\begin{thebibliography}{10}
\bibitem{analytical}
The simplified level scheme we refer to corresponds, e.g., to the case $g=0$,
where the state $|G,1\rangle$ can be neglected.
In such a three-level system, initialized to its highest state $|X^*,0\rangle$,
the following equations hold:
$ \langle \sigma^{\dagger}_r (t) \sigma_r (t) \rangle =
\Gamma_r ({\rm e}^{-\Gamma_r t} - {\rm e}^{-\Gamma_s t}) /
(\Gamma_s -\Gamma_r) $,
and
$ \langle \sigma^{\dagger}_r (t+\tau) \sigma_r (t) \rangle =
\langle \sigma^{\dagger}_r (t) \sigma_r (t) \rangle
 {\rm e}^{-\Gamma_s \tau} $
(the latter being a consequence of the quantum-regression theorem).
After integrating and performing some algebra, Eq.~\ref{distinguishability}
is derived, for $R = \Gamma_s$.
\end{thebibliography}
\end{document}